\documentclass[preprint,12pt,authoryear]{elsarticle}

\usepackage{amssymb}
\usepackage{csquotes}
\usepackage{multirow}
\usepackage[official]{eurosym}
\usepackage{pifont}
\usepackage{xtab}
\usepackage{longtable}
\usepackage{appendix}
\usepackage{pdflscape}
\usepackage{hyperref}

\journal{Journal of Systems and Software}

\begin{document}

\begin{frontmatter}



\title{Lean Internal Startups for Software Product Innovation in Large Companies: Enablers and Inhibitors\tnoteref{t1,t2}}
\tnotetext[t1]{This is the authors' version of the manuscript accepted for publication in the Journal of Systems and Software. Copyright owner's version can be accessed at \url{https://doi.org/10.1016/j.jss.2017.09.034}. \textcopyright 2018.  This manuscript version is made available under the \href{http://creativecommons.org/licenses/by-nc-nd/4.0}{CC-BY-NC-ND 4.0 license.}}
\tnotetext[t2]{Please cite as: Edison. H, Sm\o rsg\aa rd, N. M., Wang, X. and Abrahamsson, P. (2018). Lean Internal Startups for Software Product Innovation in Large Companies: Enablers and Inhibitors. \textit{Journal of Systems and Software}, 135:69--87. }


\author[unibz,ssrn]{Henry Edison\corref{cor1}} \ead{henry.edison@inf.unibz.it}
\author[ntnu]{Nina M. Sm\o rsg\aa rd} \ead{nina.m.smorsgard@gmail.com}
\author[unibz,ssrn]{Xiaofeng Wang} \ead{xiaofeng.wang@unibz.it}
\author[unij,ssrn]{Pekka Abrahamsson} \ead{pekka.abrahamsson@jyu.fi}
\cortext[cor1]{Corresponding author}

\address[unibz]{Free University of Bozen-Bolzano, Piazza Domenicani 3, Bolzano 39100, Italy}
\address[ntnu]{Norwegian University of Science and Technology, H\o gskoleringen 1, 7491 Trondheim, Norway} 
\address[unij]{Faculty of Information Technology, P.O. Box 35, FI-40014, University of Jyv\"{a}skyl\"{a}, Finland}
\address[ssrn]{Software Startup Research Network \url{https://softwarestartups.org/}}

\begin{abstract} 
\textit{Context:} Startups are disrupting traditional markets and replacing well-established actors with their innovative products.To compete in this age of disruption, large and established companies cannot rely on traditional ways of advancement, which focus on cost efficiency, lead time reduction and quality improvement. Corporate management is now looking for possibilities to innovate like startups. Along with it, the awareness and the use of the Lean startup approach have grown rapidly amongst the software startup community and large companies in recent years.

\textit{Objective:} The aim of this study is to investigate how Lean internal startup facilitates software product innovation in large companies. This study also identifies the enablers and inhibitors for Lean internal startups.

\textit{Method:} A multiple case study approach is followed in the investigation. Two software product innovation projects from two different large companies are examined, using a conceptual framework that is based on the method-in-action framework and extended with the previously developed Lean-Internal Corporate Venture model. Seven face-to-face in-depth interviews of the employees with different roles and responsibilities are conducted. The collected data is analysed through a careful coding process. Within-case analysis and cross-case comparison are applied to draw the findings from the two cases.

\textit{Results:} A generic process flow summarises the common key processes of Lean internal startups in the context of large companies. The findings suggest that an internal startup can be initiated top-down by management, or bottom-up by employees, which faces different challenges. A list of enablers and inhibitors of applying Lean startup in large companies are identified, including top management support and cross-functional team as key enablers. Both cases face different inhibitors due to the different process of inception, objective of the team and type of the product.

\textit{Conclusions:} The contribution of this study for research is threefold. First, this study is one of the first attempt to investigate the use of Lean startup approach in the context of large companies empirically. Second, the study shows the potential of the method-in-action framework to investigate the Lean startup approach in non-startup context. The third contribution is a general process of Lean internal startup and the evidence of the enablers and inhibitors of implementing it, which are both theory-informed and empirically grounded. Future studies could extend our study by addressing the limitations of the research approach undertaken in this study.
\end{abstract}


\end{frontmatter}



\section{Introduction}
\label{sec:introduction}

Today, software startups have become one of the key drivers of economy and innovation. In
2016, 550,000 new businesses or startups have been established each month in the US only
\citep{fairlie16}. Even though they are inexperienced, young and immature \citep{sutton00}, their products are disrupting traditional markets and are putting well-established actors under pressure. Uber, Spotify, and Airbnb, to name just a few, are examples of software startups that have grown rapidly. Startups offer new product, new business model, and new business value at high speed, and with cutting edge technology. They continuously talk to their potential customers to discover gaps in the existing offers, iterate, and conduct experiments to find repeatable and scalable business models. They are willing to pivot immediately if the opportunity does not prove viable. 

To compete in this age of disruption, large and established companies cannot rely on traditional ways of advancement, which focus on cost efficiency, lead time reduction or quality improvement \citep{rejeb08}. Corporate management is now looking for new ways to keep their leading positions in a fast moving market, and to innovate like startups. With greater resource in-house, they hope that they can bring innovative products with new customer values to market as startups do.

Along with it, the awareness and use of the Lean startup approach have grown rapidly amongst the software startup community in recent years. Similar to many precedent methods, the development and promotion of Lean startup have been almost entirely driven by practitioners and consultants, with little participation from the research community during the early stage of its evolution. However now it is the focus of more and more research efforts \citep{unterkalmsteiner16}.

Even though the Lean startup approach is originated in software startups, it has also gained interest from large companies as General Electric, 3M, Intuit, etc. A recent survey on 170 corporate executives reveals that 82\% of them are using some elements of Lean startup in their context \citep{kirsner16}. \cite{marijarvi16} report on the experience of large Finnish large companies in developing new software through internal startups. More and more large companies adopted the Lean startup approach, hoping that it will help them to generate successful software product innovation. 
 
\cite{ries11} argues that the core ideas behind Lean startup can offer benefits for large companies as well. If the obstacles can be minimised, the opportunities can be very beneficial to support software product innovation. Hence, evidence for understanding the enablers and inhibitors for Lean internal startups in large companies needs to be gathered. However, scientific and empirical studies regarding the leverage of the Lean startup approach for software product innovation in large software organisations are rare. Based on this observation, the main research question investigated in this study is:
\textit{How could large companies run effectively Lean internal startups for their software product innovation projects?}

To answer the main research question, we divided it into two sub-questions as follows:
\begin{itemize}
	\item RQ1: How are Lean internal startups run in large companies for their software product innovation projects?
	\item RQ2: What are the enablers and inhibitors of running Lean internal startups in large companies?
\end{itemize}

The remainder of this paper is structured as follows. Section \ref{sec:background} discusses the background and related work. Section \ref{sec:theoretical_framework} presents the theoretical frameworks used in this study, whilst Section \ref{sec:research_method} describes the research methodology employed. The key processes of Lean internal startups are reported in Section \ref{sec:findings}. Section \ref{sec:comparison} presents the enablers and inhibitors for Lean internal startups in the context of large companies, which are further discussed in Section \ref{sec:discussion}. The conclusions and future work are covered in Section \ref{sec:conclusion}.

\section{Background and Related Work}
\label{sec:background}

\subsection{Software Product Innovation}

Software product innovation is the creation and introduction of a new software product to an existing or new market \citep{lippoldt09}. The new product is developed to respond to either a technology or market opportunity \citep{krishnan01}. Newer technology is used to improve the current or to offer completely new functionalities, for example, the use of cloud computing as the online storage or the implementation of new electronic payment method. New products may be triggered by the unmet customer needs from current solutions or to address newly revealed customer needs.

In software industry, the majority of innovation could be either process or product \citep{simonetti95}. Software process innovation refers to the implementation of new processes, tools or methods to develop software, e.g., object-oriented development \citep{fichman93}, CASE (Computer-Aided Software Engineering) tools \citep{orlikowski93}, open source software \citep{feller00}, and software process improvement initiatives \citep{bygstad05}. However, the use of innovative tools or processes does not necessary lead to innovative products \citep{carlo11}.

\cite{highsmith01} claim that agile development support software process innovation by focusing on people and team. Agile seems able to prescribe on how to develop a working software faster, but is still unable to give answer what product should be developed \citep{bosch13}. Although agile also advocates to build the software iteratively, it only works when the problem is known to the stakeholders. This is not the case in product innovation, where the problem and solution are unknown.

Product innovation in software industry which is influenced either by new hardware or software development raises strategic challenges for software companies \citep{kalternecker15}. The shift from mainframe to personal computers created new market for standalone operating system. Microsoft, a new startup at that time, emerged and offered new operating system called DOS. For over a decade, the popularity of mobile devices has attracted new startups to develop various mobile apps, including new mobile operating systems, e.g., Android, iOS, etc.. Another example is the shift from proprietary software to open source software \citep{bonaccorsi06}, which allows new startups to enter a market and challenge market leaders, e.g., Linux vs. Microsoft Windows or Mozilla vs. Internet Explorer. 

In large and high-tech companies, innovative activities are performed by a specialised and dedicated entity, typically R\&D department. In R\&D, most innovations are scientific and/or technological based. The involvement of companies in R\&D activities are driven by the need to improve current process or products, researching new process or technology or specific user need. When the technology becomes more advanced and complex, R\&D are demanded to bring more innovative products. However, not all technologies produced by R\&D are inline with and directly support the business goal. These technologies are called misfit technologies \citep{anokhin11}. When this happens, the company has three options: keep scientific research, sell the technologies outside or introduce spin-off \citep{abetti02,anokhin11}.

Our previous work shows that the current research on software product innovation is scattered in different areas: early user integration, continuous experimentation, and open innovation \citep{edison16a}. Research on early user integration focuses on capturing new ideas from outside companies, i.e. users, customers, competitors etc., and turn them into real products  \citep{bailey10,blohm11,kauppinen07,gassman06}. Rather than developing new products internally, research on open innovation suggests to collaborate with external entities, e.g., through living lab. 

An emergent research area in software product innovation is startup experimentation approach, which is inspired by the Lean startup approach \citep{fagerholm14,lindgren15}. In this approach, software is developed and validated through experiments with all stakeholders. \cite{bosch12} proposes an innovation experimentation system to minimise research and development (R\&D) investment and increase customer satisfaction. In this system, R\&D runs a 2-4 week sprint based on customer feedback. However, the method is limited to SaaS (Software-as-a-Service) and embedded systems. Based on Bosch's study, \cite{fagerholm14} and \cite{lindgren15} propose a continuous experimentation system, which continuously tests the value of a product to its users. These studies emphasise more on product development itself and how to capture a product's value. An end-to-end view of software product innovation, i.e., from the generation of an innovative product idea to the realisation of its market potential, is rarely seen in the existing studies.

\subsection{Lean startup approach}
The Lean startup approach was introduced as a new way of entrepreneurship and bears the potential for product innovation in the extreme situation, where the problem and solution are unknown \citep{ries11}. Instead of emphasising a business plan, Lean startup advocates to build the product iteratively and deliver it fast to the market for early feedback. However, since customers are often unknown in the beginning, what customers could perceive as value is also unknown. Thus, entrepreneurs should ``get out of the building'' from day one to identify and validate the problem they intend to solve and discover who their customers are \citep{blank07}.

\begin{figure*}[htbp]
    \centering
    \includegraphics[width=\textwidth]{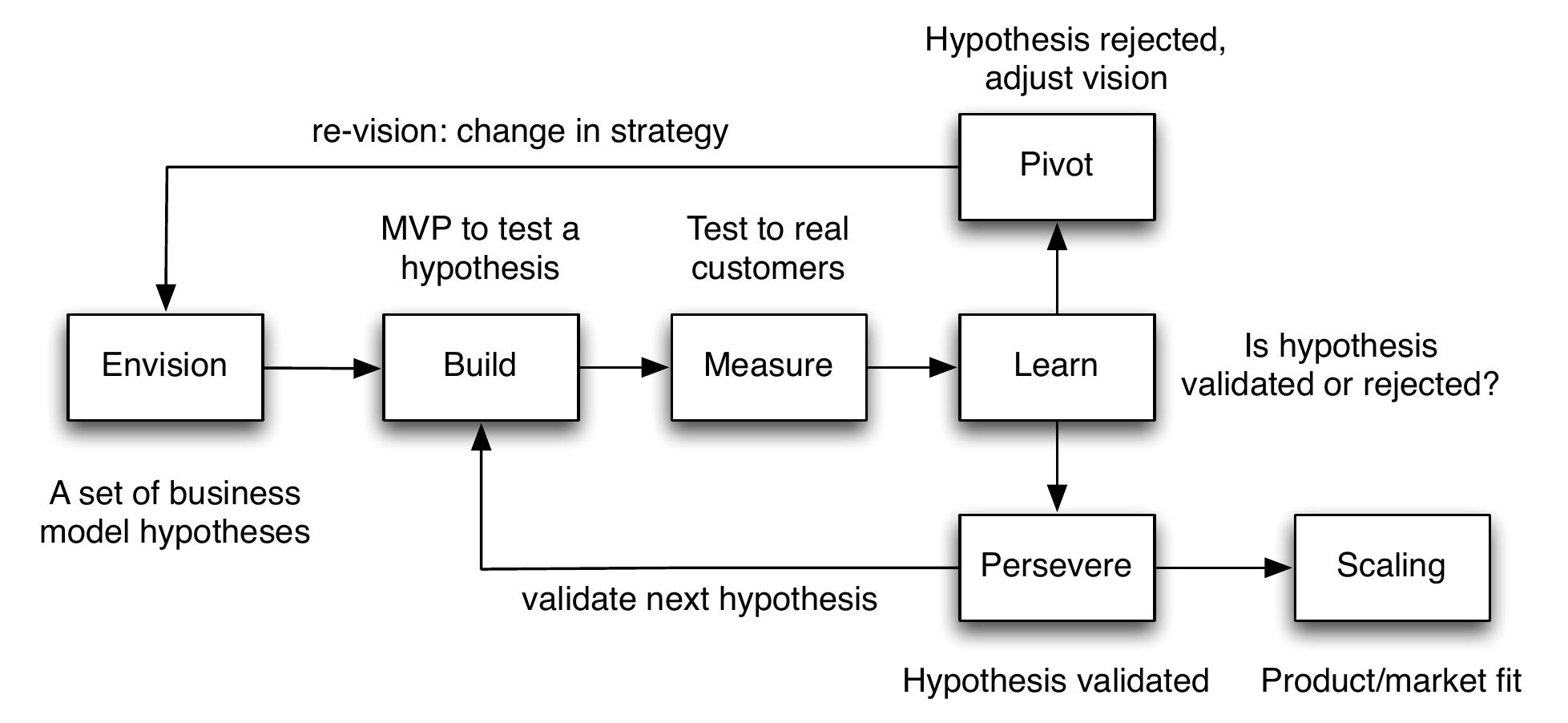}
    \caption{Lean startup process steps \citep{edison15}}
    \label{fig:lean_startup}
\end{figure*}

The Lean startup approach is built upon the Customer Development Model \citep{blank07} which consists of four steps: customer discovery, customer validation, customer creation and organisation building. The first two steps are concerned with identifying what customers value most. The last two steps aim to create a market for the product and scale the business. The model teaches to focus on and scale something that has been proven to work. Based on it, Lean startup is a structured process to validate business hypotheses through an engineering method. Fig. \ref{fig:lean_startup} presents the key processes of the Lean startup approach.

To perceive customer value, an entrepreneur starts a feedback loop (Build-Measure-Learn or B-M-L) that turns a business idea into a product.  This can be done by developing a minimum viable product (MVP) using agile methods as a tool to collect customer feedback on the product. Through the feedback, the entrepreneur learns whether to persevere on the proposed business idea or to pivot to a new direction, or to perish -- renounce the business and the product \citep{eisenmann13,ries11}. The key practices of Lean start-up are summarised in Table \ref{tab:key_practices}.

\begin{table*}[htbp]
    \caption{Key practices of Lean startup approach (adapted from \citep{ries11})}
    \label{tab:key_practices}
    \begin{tabular}{|p{3cm}|p{9.7cm}|}
    \hline
    Key Practice & Description\\ \hline
    Get-out-of-building & Confirm through face-to-face interaction with customers specifically what the problem is and whether it is worth solving. The purpose of early contact with customer is to understand the potential customers and their real problems. \\ \hline
    MVP & To validate the leap-of-faith assumptions, a version of product with minimum amount of effort should be released as quickly as possible. If MVP seems to have dangerous branding risk, launch MVP under different brand name. \\  \hline
    B-M-L loop & A feedback loop, which in order to turn ideas into products, measures how customers respond and learns whether to pivot or persevere. \\ \hline
    Use actionable metrics	& Metrics that demonstrate clear cause and effect to evaluate the progress.\\ \hline
    Small batches & Engineers and designers work side by side on one feature at a time. Whenever that feature is ready to be tested with customers, they release to a small number of people. \\ \hline
    Pivot & Change in course or strategy. There are 10 types of pivot proposed in \citep{ries11}: zoom-in pivot, zoom-out pivot, customer segment pivot, customer need pivot, platform pivot, business architecture pivot, value capture pivot, engine of growth pivot, channel pivot, technology pivot. \cite{Bajwa2016} identified several new pivot types, including side-project pivot and complete pivot.\\ \hline
    Continuous deployment	 & The code written for an application is immediately deployed into production.\\
       \hline
   \end{tabular} 
\end{table*}

Current research on the Lean startup approach is centred on applying the method in a standalone startup context to develop new product, e.g., \citep{haniotis11,may12,efeoglu14}. Very few peer-reviewed studies investigate how the Lean startup approach supports software product innovation in large companies. Our previous study based on a single case study finds that internal startups share the same characteristics as standalone startups, which is aiming at product innovation \citep{edison15a}. In this study we extend our previous research by investigating two internal startups in two different companies.

\subsection{Internal Corporate Venture (ICV)}
ICVs are corporate entrepreneurial efforts that originate within a corporation and are intended from inception as new business for the corporation \citep{kuratko09}. ICVs operate as semi-autonomous corporate startups \citep{simon99} or innovation hubs \citep{ohare08} or internal startups \citep{huumo15}. The introduction of a new internal venture may be the consequence of following or leading to product or market innovation \citep{sharma99,block93}. The degree of newness is defined by being new in the world and new in the industry \citep{kuratko09}. New business can be established as an instrument to pursue incremental innovation (a new product in a current market or a new market for a current product) or radical innovation (a new product for a new market). Innovation is generated through a separate and dedicated entity which is operated within an established company, using resources that are solely under the control of the company \citep{roberts85,narayanan09}.

In the context of large software companies, a study by \cite{raatikainen16} investigates how internal startups are used for new product development. The study finds that in each phase of a new product development life-cycle, companies can apply different structures, e.g., internal startup, company subsidiary, incubating, etc.. Another study by \cite{selig16} investigates the role of corporate entrepreneurs in internal startup. The study finds that corporate entrepreneurs share the same characteristics as independent entrepreneurs. To further pursue innovation, corporate entrepreneurs need a guarantee of minimum salary and autonomy to experiment.
\section{Conceptual Framework for Lean Internal Startups}
\label{sec:theoretical_framework}
Since the main focus of our study is the application of the Lean startup approach in software product innovation projects in large companies, to make better sense of the research phenomenon, we draw upon the method-in-action framework proposed by \cite{fitzgerald02}. It is a conceptual framework to investigate the use of a method in the complex environment of software and system development. The framework has been widely used in the information system (IS) literature, e.g., \cite{backlund02,mihailescu06,oneill11}. It does not prescribe detailed and specific actions to be used. It allows us to reflect on the IS development as rich and complex phenomenon influenced by the components and their interactions \citep{oneill11}.

\begin{figure*}[htbp]
    \centering
    \includegraphics[width=0.83\textwidth]{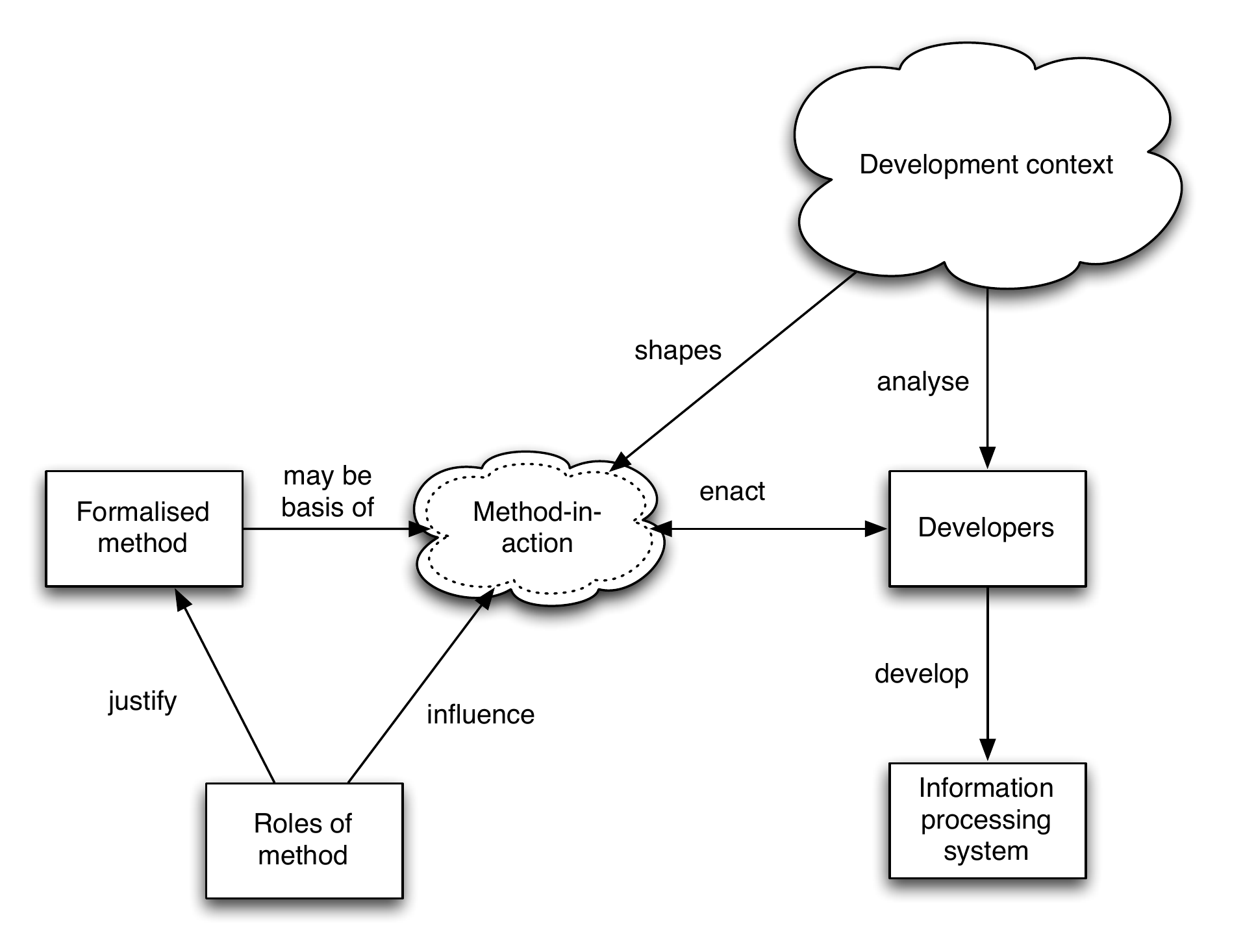}
    \caption{Method-in-action framework \citep{fitzgerald02}}
    \label{fig:mia}
\end{figure*}

Fig. \ref{fig:mia} is an overview of the method-in-action framework. The method recognises different components that affect the practice of the method. \textit{Formalised method} refers to a commercial or in-house method the usage of which is formally documented. A formalised method may serve as a reference or guide for the usage of the method in action. Developers uniquely enact a  \textit{method in action}, which is reflected by the cloudy outlines. Regarding the \textit{roles of method}, there are two categories of roles that mediate how methods work: rational and political roles. A set of rational roles is overtly used to justify part of the conceptual basis and the rationale underlying the use of the method. In contrast, a set of political roles is covert in nature and influences the derivation of the method-in-action. \textit{Developers} play a central role in the framework, since they refer to the stakeholders, e.g. programmers, designers, etc., who analyse the \textit{development context} to develop an \textit{information processing system}. Such a system can be described, identified and understood in different families of systems, which affect the development process.

\begin{figure*}[htbp]
	\centering
	\includegraphics[width=0.83\textwidth]{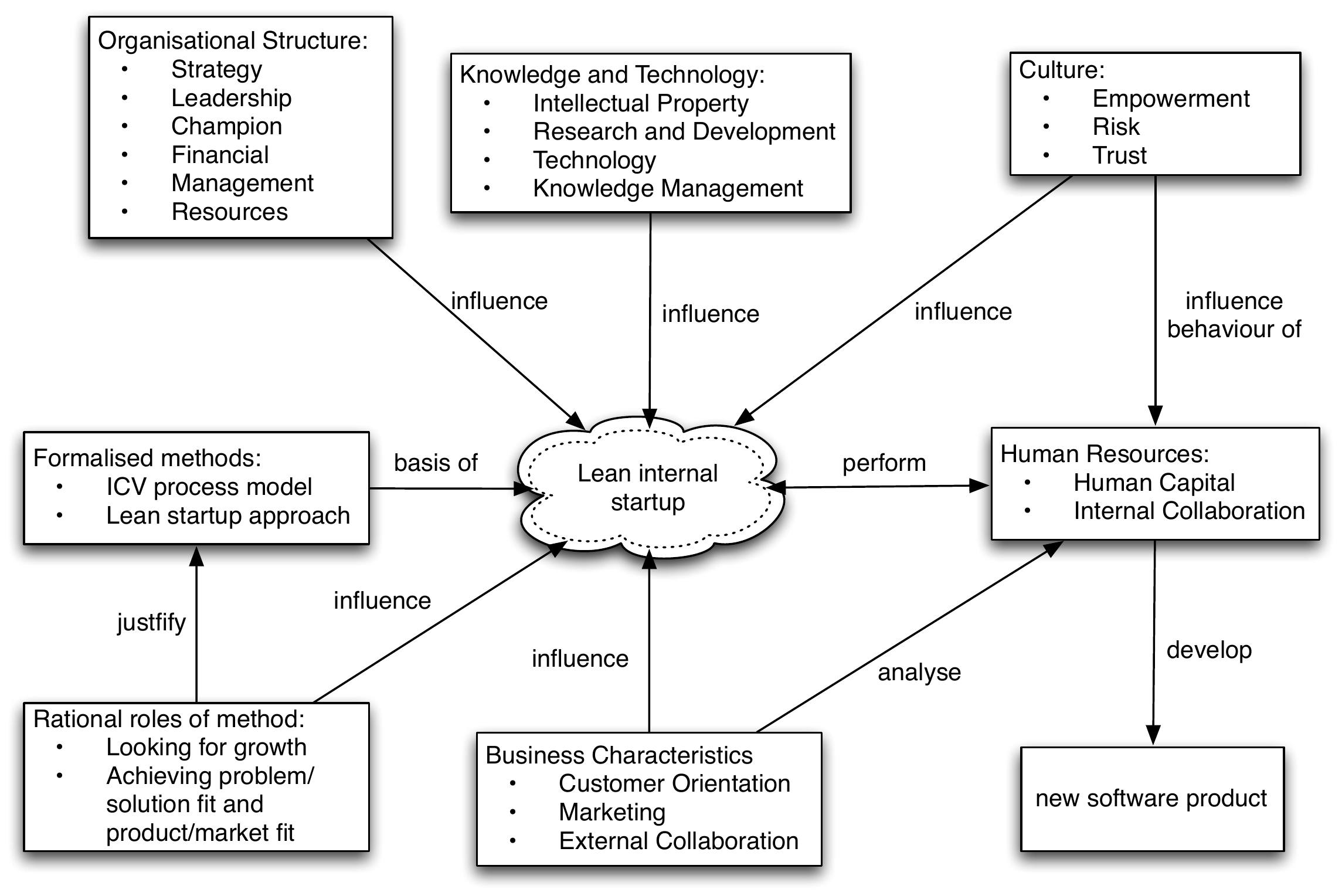}
	\caption{Conceptual framework for Lean internal startup (adapted from \cite{fitzgerald02})}
	\label{fig:li}
\end{figure*}

We adapted the method-in-action framework to study Lean startup in action in the context of large companies. Fig. \ref{fig:li} illustrates the conceptual framework for Lean internal startup. Research has shown that the Lean startup approach is originated in standalone software startups. Hence, its adoption and interpretation in large companies are influenced by a number of determinants related to their context: organisational structure, knowledge and technology, culture, human resources, business characteristics. In our conceptual framework, the development context and developers are replaced with these determinants. The determinants for innovation success have been proposed in \cite{edison13}, who performed a literature review of the empirical literature in order to identify success factors of new product development in the software context.  The determinants that are relevant to this study are introduced when each framework component is described in the following subsections. The complete list of determinants of innovation can be found in \cite{ali11}. 

\subsection{Formalised Methods}
Originally, the Lean startup approach was designed to manage startups in order to speed up the product/market fit \citep{ries11}. The approach helps entrepreneurs to find out whether a product should be built. \cite{ries11} argued that large companies can also benefit from practising Lean startup approach. However, a startup is not a small version of a corporation and corporation is not a large version of a startup. Since large companies rely on a management structure, they tend to be bureaucratic. Any attempt to change the stability will be considered a violation of certain territorial rights \citep{shepard67,ahmed98,gorschek10}. As discussed in Section \ref{sec:background}, ICV is deemed as an important avenue for nurturing innovation and entrepreneurship in large companies. 

The Lean-ICV framework \citep{edison15} was an attempt to integrate the Lean startup approach into the ICV process model \citep{burgelman83}. Both are complementary to each other. While the ICV process model describes the process using the company perspective, the Lean startup approach describes the activities from the team perspective. Table \ref{tab:lean_icv} presents the Lean-ICV framework. This model acknowledges the dynamics of both the process at the innovation team level to achieve the product/market fit and at the corporation level, which is aimed at keeping the initiative within the company's boundary.  
    
 \subsection{Lean Internal Startup}
The method-in-action framework suggests that formalised methods are used in ways different than intended by their creators. In the conceptual framework, the method-in-action is replaced with the term Lean internal startup, which signifies the Lean startup approach actually being used in the context of large companies. The cloud shape used to depict this component reflects the somewhat undefined nature of its content. As with the method-in-action framework, this central component is meant to reflect the actual internal startup activities performed in a large company to develop new software products. 

\subsection{Roles of Method}
The roles of method that emphasised in this study are the rational intellectual ones: looking for growth and achieving problem/solution fit and product/marketing fit. Academic research into new product development reveals that successful product innovation is vital to many established and large companies, but many innovators are failed to develop a product/market fit \citep{dougherty92}. While developing a new product, each unit or department in large companies is like a different world of thought and focuses on different aspects of technology-market knowledge. Instead of being coordinated, these worlds of thought are separated in terms of their organisational routines, which impedes joint learning.

In terms of the level of integration with the core organisation, different organisational structures have been proposed in the literature \citep{ohare08}. On the one side is the R\&D unit, which is responsible for leveraging resources in the company in order to support the main core of the business. Hence, R\&D units cannot go too far beyond the core competences of the company. On the other side are company spin-offs. Spin-offs enjoy the freedom and autonomy of being able to develop their own process and culture. Somewhere in the middle of this spectrum are ICVs.

\begin{landscape}
\begin{table*}[htbp]
    \caption{Lean-ICV Framework \citep{edison15}}
    \label{tab:lean_icv}
    \begin{tabular}{|p{2.5cm}|p{3.5cm}|p{13cm}|}
    \hline
    Major process & Description & Key activities \\ \hline
    Envisioning & Setting up the vision & \textit{Vision}: Adopting the Lean startup approach, the innovation initiative starts with envisioning, where intrapreneurs set the visions for a new product and translates them into hypotheses. To do this, they need two things: \textit{authorisation} from corporate management and \textit{coaching} from NVD (New Venture Devision) management on how to turn the vision into successful business. \\ \hline
    Steering & Validating hypotheses & The initiative needs a product champion to get further resources. Once it gets approval from top management, the \textit{build-measure-learn loop} takes place to validate all hypotheses. Based on this learning, intrapreneurs have two options: pivot or persevere. When all the hypotheses are valid, then it is the time to integrate the new business into the company's portfolio. Since the internal startup uses the parent company's resources, during the development, corporate management \textit{monitors} the progress of initiatives. \\ \hline
    Accelerating & Leveraging the new product into the strategic context &  In this process, the intrapreneurs are finding a way to \textit{scale} the business. The business model hypothesis have been proven to generate financial benefits. To control internal startup initiatives in the company, corporate management uses \emph{selection} mechanism. Only the initiatives that have greater potential impact get continuous support. In the rationalising process, the intrapreneurs must convince corporate management to change the strategy to accommodate the new business. In parallel, the NVD management plays an important role in \textit{delineating} the new business in the current strategy. Therefore, organisational championship is needed to continuously communicate with corporate management regarding the development of new business idea. \\ \hline
               \end{tabular} 
\end{table*}
\end{landscape}

As discussed in Section \ref{sec:background}, to help achieve product/market fit, \cite{ries11} introduced Lean startup approach. The Lean startup approach is a structured process for validating business hypotheses through an engineering method. A product/market fit is defined as (1) the customer is willing to pay for the product, (2) there is an economical viable way to acquire customers, and (3) the market is large enough for the business \citep{cooper10}.

\subsection{Organisational Structure}
The fourth component of the framework is the organisational structure. An organisational structure defines how activities are controlled and coordinated in order to achieve the organisational goals \citep{koberg96,menguc00,chang08}. The relevance within the conceptual framework is that the organisational structure plays an important role as it may either hinder or support innovation in place. Unlike startups, large companies already have ongoing big businesses, as well as having their shares of markets and customers. To manage these businesses, large companies heavily rely on bureaucracy, standardisation and formalisation. Bureaucracy employs institutionalised rules, policies and routines that define how tasks are to be accomplished. Standardisation governs how employees interact, and how decision making is achieved. Moreover, employees already have specific formal jobs and responsibilities. 

From an innovation perspective, every large company faces the dilemma of having to serve the existing market, and at the same time, striving for growth \citep{ford10}. This rigid and formal nature of the organisational structures impedes creativity, risk-taking, exploration and experimentation. Implementing out-of-the-box types of thinking and behaviour is not allowed in such an environment, since everything must adhere to predefined rules, practices and routines. On the other hand, large companies need to remain stable in a dynamic and disruptive environment while also creating new business opportunities. 

The Lean internal startup approach is an attempt to allow large companies to innovate like startups. As described in Section \ref{sec:background}, it requires different treatment than the existing business. As in other innovation initiatives, organisational characteristics will influence how Lean startup is applied in large companies. In this study, the effect of the organisational structure on Lean internal startup is examined by looking at strategy, leadership, champion, financial situation, management, and organisation resources which show positive contribution to innovation success \citep{edison13}.

\subsection{Knowledge and Technology}
In large companies and in high-tech industry, innovative activities are performed by a specialised and dedicated entity, typically the R\&D department. In R\&D, most innovations are scientific and/or technology based. The involvement of companies in R\&D activities is driven by the need to improve current processes or products, by researching new processes or technologies, or specific user needs. In fact, economies of scale in R\&D, risk diversification and access to greater financial success are the main benefits that large companies get from generating radical innovation \citep{ford10}.

When the technology becomes more advanced and complex, R\&D units are demanded to bring forth more innovative products. However, not all technologies produced by R\&D are inline with and directly support the business goal. These technologies are called misfit technologies \citep{anokhin11}. When this happens, the company has three options: keep on doing scientific research, sell the technologies outside or start a spin-off \citep{anokhin11,abetti02}. The successful application of technology-based innovation is determined by several key factors, including intellectual property, research and development activity, existing technology available and knowledge management \citep{edison13}. 

\subsection{Culture}
The sixth component of the framework is culture. The common perception of culture relates to the values and beliefs shared by the employees in a company \citep{ahmed98,martins03}. Organisational culture significantly influences employees' behaviour and attitude to perform beyond formal control systems, procedures and authority \citep{oreilly91}. Organisational culture supports the development of creative solutions, and thus encourages innovative ways of representing problems and finding solutions \citep{martins03}.

The key culture-related determinants of innovation are empowerment, trust and risk \citep{edison13}. The literature on corporate entrepreneurship also suggests that companies need to nurture corporate entrepreneurs by having an innovative and entrepreneurial culture \citep{morse86,kuratko14}. At the strategic level, corporate management should recognise that innovation entails risk and employees may work in unpredictable ways of doing things \citep{myers84}. Some are great visionaries and willing to pursue them but some are very effective to imitate an idea and adapt it to a new setting. Some are very creative to seek a gap in the current market and fill it \citep{myers84}. Hence, management must support, facilitate and encourage entrepreneurial behaviour \citep{kuratko14}. At the tactical level, companies should empower employees and trust them if they want to engage in any innovation initiatives. Failure must be considered as part of the learning process.  
 
\subsection{Business Characteristics}
The seventh component of the framework is business characteristics, which includes customer orientation, marketing and collaboration with external suppliers \citep{edison13}. Customer orientation is one of the important factors that significantly affects a company's capability to innovate \citep{akman08}. It describes the company's behaviour to understand and create high value for the fulfilment of their customers' needs. The authors argue that by focusing more on customers, companies will be able to improve their product innovation since customers' needs and wants are the sources of innovative ideas. The same finding is also mentioned by \citep{paladino07,voss85}. Therefore, managers should look for a new strategy to fulfil the market needs although this is difficult to achieve. Companies should not focus on current needs but also on future needs. This can be done when an organisation maintains a good relationship with customers.

Among the six sets of general new product development \citep{song98}, product marketing or commercialisation is one of the key activities that showed the most significant differences between success and failures \citep{cooper86}. It includes coordinating, implementing, and monitoring the new product launch. Product commercialisation activities can also be used to gain new knowledge and information directly about the market and customers. This new knowledge becomes a new source for innovation \citep{kline85}.

Studies shows that collaborative working with external suppliers make a significant contribution to the product innovation process e.g. the use of guest engineers, joint project with third parties \citep{adams06}. Companies can get significant benefits by involving suppliers in the early stages of the product innovation process \citep{huang00}.

\subsection{Human Resources}
In the method-in-action framework, developers have a central role, because they develop the system, not the method \citep{fitzgerald97}. The term ``developers'' refers to multiple stakeholders: system users, analysts, designers, programmers, clients and problem owners. As described in Section \ref{sec:background}, the Lean startup approach is about building a sustainable business model rather than a new product. Therefore, in this conceptual framework, the term ``developers" does not only refer to the people who are responsible for product development but also to the people who are responsible for business development, e.g., marketing, legal, etc..

From an innovation perspective, two key factors of successful product innovation are: (1) the quality of the people who are directly involved in the product development, and (2) the collaboration between the team members and other people in the company \citep{edison13}. Good and integrated collaboration and coordination among all departments can promote an effective knowledge transfer within the company. It allows sharing of innovative ideas among the employees and transforming them into innovative outcome.
\section{Research Approach}
\label{sec:research_method}
The research questions and the conceptual framework required that we examined the Lean startup approach in the context of organisations that adopted it. Therefore, case study is a suitable approach to better understand this phenomenon, investigating it in its real-life setting. In addition, the case study approach is also beneficial where control over behaviour is not required or possible \citep{yin09}. This study uses a multiple-case design to allow a cross-case pattern search. 

Following the case study guidelines proposed by \cite{runeson12}, we consider the present study an exploratory case study as it is not concerned with theory or hypothesis testing. In addition, this study intends to examine different software product innovation processes and to identify the organising patterns behind them. The research results would be more convincing if similar findings emerge in different cases and evidence is built up through a family of cases.

Given the research questions of the study, the level of inquiry is at the team level. Hence, it seems appropriate to take an internal startup team as a case. Since the focus is on the software product innovation process, the unit of analysis is the software product innovation process. The process is employed by the team; therefore, the unit of analysis is at the same level as the case. In this study, the case selection criteria are: 
\begin{itemize}
	\item Each case comes from a large company. Large companies are defined by the following criteria \citep{eu15}: (1) staff headcount: employ $>$ 250 persons, and (2) annual turnover: $>$ \euro 50 million, or balance sheet total: $>$ \euro 43 million.
	\item Each company has established a dedicated team that is responsible from ideation to commercialisation of a new software.
	\item The software is targeted at external users or customers.
	\item The software falls outside of the current main product line, thus representing a more radical product innovation.
\end{itemize}

\subsection{Case companies and products}
\label{sec:case_companies}
This study investigated two cases from two different companies. At the request of the companies, they and the product innovation projects under the study will remain anonymous throughout the paper. Both companies were identified with the help of researchers who are part of the Software Startup Research Network (SSRN) \footnote{https://softwarestartups.org/}. The profiles of the two case companies and the teams are presented in Table \ref{tab:companies_profile} and Table \ref{tab:teams_profile}.

\begin{table}[htbp]
    \caption{Profile of the two case companies}
    \label{tab:companies_profile}
    \begin{tabular}{|p{3cm}|p{4.7cm}|p{4.7cm}|}
    \hline 
     Case name& FastCaf\'{e} & SeeSay \\ \hline
     Company name & CallBook & CallTech \\ \hline
     Company background & A print directory publishing company & A telecommunication company \\ \hline
     Total \# of employees & $>$ 2,000 & $>$35,000 \\ \hline
     Total revenue & \euro 352 million (2015) & \euro 14,000 million (2015)\\
       \hline 
   \end{tabular} 
\end{table}

\begin{table}[htbp]
    \caption{Profile of the two case teams}
    \label{tab:teams_profile}
    \begin{tabular}{|p{3cm}|p{4.7cm}|p{4.7cm}|}
    \hline 
     Case name& FastCaf\'{e} & SeeSay \\ \hline
     Team size & 7 -- 15 & 4 -- 20 \\\hline
     Team composition & Team lead, User Experience (UX) designer, 3 developers, 2 part-time members at start; now has 15 members & Team lead, 3 developers (interns) at the start; now has 20 members \\\hline
     Product type & Online pre-payment platform & Audio \& video conversation platform \\\hline
     Development timeframe & 2014 -- now & 2013 -- now \\\hline
     Current stage & Accelerating & Steering \\
       \hline 
   \end{tabular} 
\end{table}

\subsubsection{CallBook}
CallBook is one of the leading marketing companies and one of the largest print directory publishers. The initiative for new product development has been part of the company strategy to shift from a print directory business to a digital business. Revenues from its print-based business are declining at an average of 15\% a year. In 2012, CallBook invested in innovation skills by bringing in a design consultant, to kickstart a design thinking capability for its new product. The management wanted to find a way to diversify the product portfolio. As the results of this initiative, two product innovation projects were established to start the new product development. In the middle of development, new management came in and evaluated the ongoing projects. The new management found that the first project was 7--8 weeks behind schedule. As a consequence, the first project was terminated. 

The second product innovation project was FastCaf\'{e}. There was no formal internal human resources (HR) process to recruit the team members. If the employees were interested, they only needed to talk to their managers about this opportunity. From the many employees who showed an interest in joining the team, 6--7 were selected based on their skills and attitudes. The team members were individuals who had deep knowledge in one or two areas but still had adequate knowledge across all areas more broadly so that they were able to interweave with other disciplines to fill in any gaps.

\subsubsection{CallTech}
CallTech is one of the leading telecommunication companies. It is considered a hierarchical and bureaucratic organisation by the interviewees. Traditionally, as a telco, CallTech provides a good infrastructure and technology for telecommunication networks, including Internet connection. CallTech was looking for product innovation beyond the existing technology and launched an in-house intrapreneurship initiative. SeeSay was born internally in the intrapreneurship program of CallTech and the initiative was taken by one of the employees, who has now become the vice president of SeeSay.

The project started in 2010 as a summer project done by three internship students. When the internship ended, the project was continued by a team of engineers. Today, the project is scaled up to 20 employees with full-time responsibilities.

In 2016, CallTech launched Flash, a new innovation program that allows employees to develop new product ideas into testable prototypes. Successful pilots are then given resources to be developed into products and the access to the market in which CallTech operates is enabled. Today, SeeSay is operating under the Flash program.

\subsection{Data Collection and Analysis}
\label{sec:collection_analysis}
In this study, the conceptual framework serves as the theoretical lens for the investigation of the case, acting as a sensitising and sense-making device that guides the data collection and analysis processes. It was used to frame the interview questions and enabled a holistic understanding of the dynamics between the internal startup and other entities within the company. The formalised methods component of conceptual framework was used to answer RQ1, whilst the other components of the framework were used to guide the analysis for RQ2.

Semi-structured interviews were used as the primary data collection method. To better understand the phenomenon, several members from each company were interviewed. The background information of the interviewees is presented in Table \ref{tab:interviewees}. The interviewees were selected based on their involvement in the development and their availability in the interview process. In the CallBook case, the first author led the interview process whilst in the CallTech case, the second author guided the interview process.

Two pilot interviews were conducted in July and August 2014. The pilot interviews were intended to evaluate whether the questions were interpreted in the same way by the interviewer and the interviewee. Both interviews were recorded, transcribed, and analysed to clarify the research questions and to test the initial conceptual framework. Both interviewees were responsible for internal product innovation initiatives in their companies.

\begin{table*}[htbp]
   \centering
    \caption{Background information of the interviewees}
    \label{tab:interviewees}
    \begin{tabular}{|p{2cm}|p{3.5cm}|p{2cm}|p{4.5cm}|}
    \hline
    Company  & Role & Years of experience & Responsibility\\ \hline
    \multirow{2}{*}{\parbox{1.2cm}{CallBook}} & Team Lead & $>$ 10  & Product manager, leading innovation project\\ \cline{2-4}
      & UX (User eXperience) Designer Lead & $>$ 8  & Understanding customer needs and translating them into usable features.\\ \hline
     \multirow{2}{*}{CallTech} & Product Development Manager & $>$ 5  & Co-founder, Product development manager \\ \cline{2-4}
     & Team Lead & $>$ 6 & Founder and technical lead of SeeSay\\ \cline{2-4}
     & Chief Innovation Officer (CIO) & $>$ 4  & Responsible for all product innovation agenda in CallTech\\ \cline{2-4}
     & Product Manager & $<$ 2  & Product manager for SeeSay \\
       \hline 
   \end{tabular} 
\end{table*}

The data collection was initiated in May 2016 and the follow-up interview sessions were conducted when clarification and more information needed to be obtained. Seven interviews were done in three rounds. Table \ref{tab:data_collection} shows the detailed arrangement of the data collection in each round. Most of the interviews were done in the interviewees' offices, but some of them were done through Skype due to the geographical constraints. Each interview lasted between one and two hours, and was recorded. All interviews were transcribed verbatim. Notes were taken during the interviews. To achieve data triangulation, other supporting materials, such as company presentations, white papers, were also collected to supplement the interview data. We also looked at the newspapers, magazines and other published materials available to balance the information we got from the interviewees.

\begin{table*}[htbp]
    \caption{Data collection in the two cases}
    \label{tab:data_collection}
    \begin{tabular}{|p{1.8cm}|p{3.4cm}|p{3.4cm}|p{3.4cm}|}
    \hline
    Company & \multicolumn{3}{c|}{Data collection}\\ \hline
    \multirow{4}{*}{\parbox{1.2cm}{CallBook}}  & First round (Visit, 01.05.2016) & Second round (Skype call, 01.06.2016) & Third round (Skype call, 21.09.2016)\\ \cline{2-4}
   & \multicolumn{3}{c|}{\textbf{FastCaf\'{e}}} \\ \cline{2-4}
      & 1 individual interview: UX Designer  & 1 individual interview: Team Lead & 1 individual interview: UX Designer \\ \cline{2-4}
      & Documentation review & & \\ \hline
     \multirow{6}{*}{CallTech} & First round (Visit, 20.07.2016) & Second round (Visit, 14.09.2016) & Third round (Visit, 03.10.2016)\\ \cline{2-4}     
     & \multicolumn{3}{c|}{\textbf{SeeSay}} \\ \cline{2-4}
     & 1 individual interview: Co-Founder & 1 individual interview: Team Lead & \\ \cline{2-4}
     & Documentation review & Documentation review & \\ \cline{2-4}
     & \multicolumn{3}{c|}{\textbf{Senior Management}} \\ \cline{2-4}
     & & 1 individual interview: VP Product Management and Innovation & 1 individual interview: Chief Innovation Officer\\ \hline 
     & Total number of interviews: 2 & Total number of interviews: 3 & Total number of interview: 2 \\ \hline
   \end{tabular} 
\end{table*}

The interview data was analysed iteratively, following the three types of coding process \citep{yates04,saldana12}: (1) open coding, which is done by scrutinising all the sources very closely, line by line; (2) axial coding, where the analysis revolves around one category at a time; and (3) selective coding, which is systematic coding for the core category. The conceptual framework and its main components as discussed in Section \ref{sec:theoretical_framework} provided the seed categories for the coding process. The focus of the analysis was on the innovation activities and their impacts on the internal startup team and the company. The documents obtained from the interviewees and the field notes were also included in the coding steps, which allowed us to triangulate the interview data.

The first and second authors were responsible for analysing the data. Before the actual analysis process took place, the analysts conducted a meeting to agree on the coding process. Follow-up discussions were done if needed to ensure that the data from both cases were treated equally. At the end of the analysis process, the results were exchanged among the analysts to discuss and get feedback about the results and the coding process. The examples of coding process are presented in Table \ref{tab:coding_process}.
  
  \begin{table*}[htbp]
   \centering
    \caption{Examples of coding process}
    \label{tab:coding_process}
    \begin{tabular}{|p{4.8cm}|p{2.3cm}|p{2.3cm}|p{2.5cm}|}
    \hline
    Quote & Open code & Axial code & Selective coding\\ \hline
    \multicolumn{4}{|c|}{RQ1}\\ \hline
   ``We meet with CEO every fortnight without fail and we present our findings to directly to him.'' & Constantly  reporting to the CEO & Progress monitoring &  \multirow{3}{*}{\parbox{3cm}{Monitoring (Management)}} \\ \cline{1-3}
   ``At the meeting we would agree what target we would like to hit for the next six weeks, and then he would say you can keep go.'' & Continuous communication with the CEO & Getting approval from the management & \\ \hline 
    \multicolumn{4}{|c|}{RQ2}\\ \hline
    ``We were given one of the best things that happen to us, it was CEO, who was incredibly supportive and one of the things that he said to us early was I give you permission to do things differently.'' & Support from the CEO  & Leadership commitment & Leadership \\ \hline
    ``It wasn't easy in the beginning. It's not easy to begin something in a large company, and then get resources in a very early phase.'' & Difficulties to get resources in the early phase & Securing resources & Financial \\ \hline
   \end{tabular} 
\end{table*}

The further analysis of the two cases followed two steps. Firstly, each case was analysed as a stand-alone entity, or what \cite{yin09} refers to as within-case analysis€™. Then cross-case comparison was conducted to detect common patterns between the two cases, to consolidate the case study findings.
\section{Lean Internal Startup Processes}
\label{sec:findings}

In this section, we present the Lean internal startup processes of the two cases studied. Each case is described in detail. 

\subsection{FastCaf\'{e}}
\label{sec:case_a}

\begin{figure*}[htbp]
    \centering
    \includegraphics[width=\textwidth]{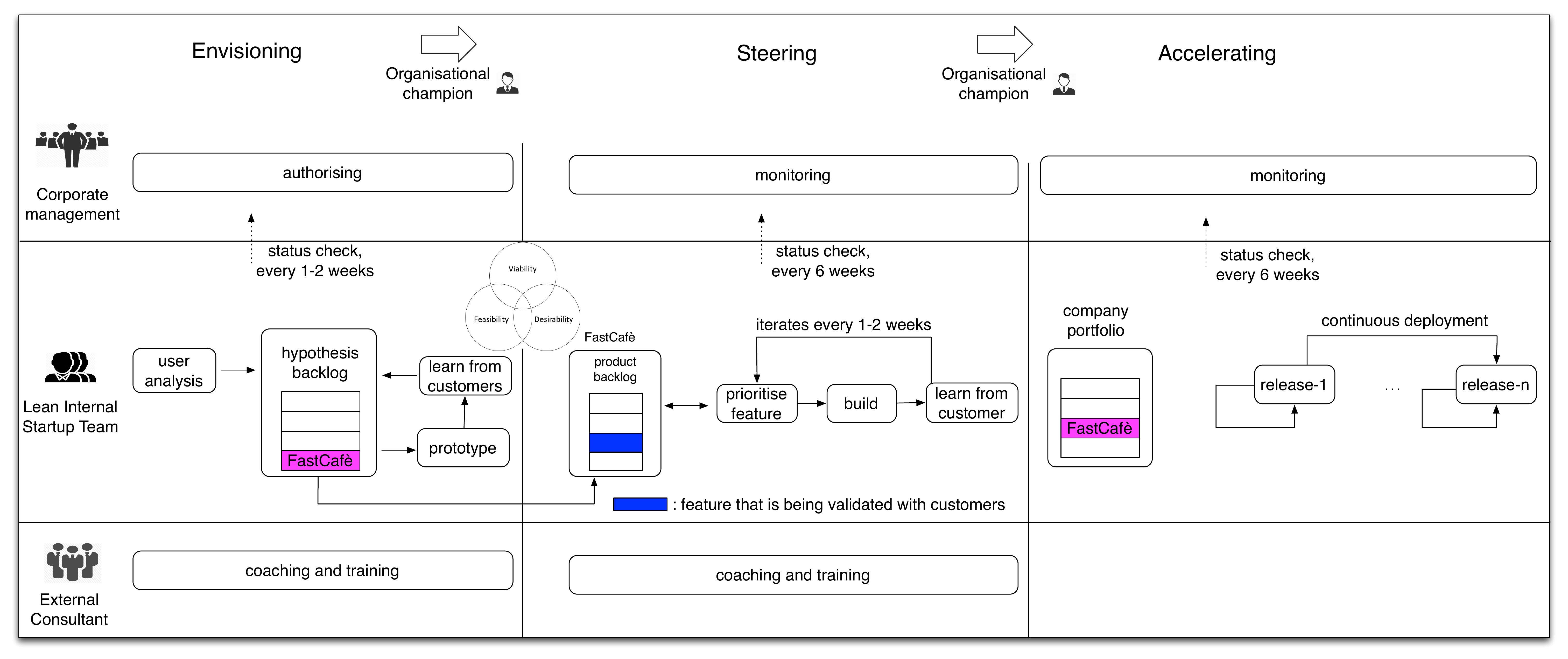}
    \caption{The Lean internal startup process of FastCaf\'{e}}
    \label{fig:case_A}
\end{figure*}

The internal startup initiative began in July 2013. All the team members were recruited internally. At the beginning, the team had seven members including developers, UX designers and team lead. As the product is becoming more mature, the team is growing in terms of size. Currently the product is in the business scaling phase and the team has more than 15 members. Fig. \ref{fig:case_A} illustrates the key process of the Lean startup approach in CallBook. The key metrics that were collected during each phase are presented in Table \ref{tab:measures}.

\subsubsection{Envisioning Phase}
\label{sec:exploration}
The initiative for new product development was part of the company strategy to diversify their product portfolio due to declining revenues in 2012. The company wanted to have a new product development team that was looking beyond the core business. An external consultant was hired to train the team in developing new products.
\begin{displayquote}
\textit{``Our top management] hired IDEO, [a] global design consulting and agency, ... and they took a group of us and trained us in [the] Design Thinking method and took [us] away from the core business of the day-to-day activities and retrained us in a way that can very quickly get a product to market and do it exhaustively quickly and eliminate a lot of barriers that occur both culturally, politically, technically in a large organisation.''} -- UX Designer Lead
\end{displayquote}

To find an idea for a new product, the team did quantitative and qualitative research on existing solutions in the market and potential users. The research on existing solutions aimed to identify their strengths and the weaknesses as well as their targeted market, and look for inspiration for a new product. The research on potential users emphasised on their behaviour set. The team went out in the streets to observe and interview people about their opinion on an idea.
\begin{displayquote}
\textit{``We went out and we were asking people ... in a way that it was very informal and friendly. One of us brought a paper prototype and a frame board with sticky notes or little stickers and [asked] `Tell us where is your favourite place to go for coffee and use the map and sticker and why is that?'''} -- UX Designer Lead
\end{displayquote}

The key information collected from the research was then discussed and synthesised by all the team members.
\begin{displayquote}
\textit{``We sit down and we have a white board and sticky notes and ... put down on the post-it notes everything that we experienced and thought about at that day and [the] team starts to discuss together and ... Then we have little red sticky dots and we go, what are the things we want to pursue next, what do we want to hypothesise on and test on and we vote on that, very democratically, and once we picked up 1, 2, 3 ideas and we write how do we design a measurable experiments for that.''} -- UX Designer Lead
\end{displayquote}

The first MVP of FastCaf\'{e} was using short message service (SMS) where people could send an SMS to order a coffee. The solution was tested with the people in the company to see how it could work. The team got around 20 orders in one week. To test with a broader audience, the second MVP was developed, an HTML--based mobile application using which customers could select the coffee, the caf\'{e} and the pick-up time. Once the order was received, the team would go to the caf\'{e} and place the actual order. As the customers came and picked up the coffee, the team interviewed them about their experience. The MVP was tested for three days. The team generated 700 dollars worth of orders and collaborated with 4 caf\'{e}s. 

\begin{table}[htbp]
    \caption{Measures collected by FastCaf\'{e}}
    \label{tab:measures}
    \begin{tabular}{|p{2.5cm}|p{5.5cm}|p{4.5cm}|}
    \hline 
    Phase & Measure & to Learn about  \\
    \hline
    \multirow{ 2}{*}{Envisioning} & The most favourite coffee & The design interface of the app\\ \cline{2-3}
    & Number of orders & Financial viability \\ \hline 
   \multirow{ 2}{*}{Steering} &  Order per day, order per week, most favourite venues & Consumption behaviour \\ \cline{2-3}
     &  Number of revenues, number of orders &  Business model \\ \hline
   Accelerating &  Number of caf\'{e}s  & Revenue growth \\
   \hline 
   \end{tabular} 
\end{table}

In the case of FastCaf\'{e}, the CEO took the organisational championship. He was involved since the inception of FastCaf\'{e} in the company by hiring a consultant company to give a training on design thinking. He was also the one who backed up the team when a new management came in.

During this phase, the team met with the CEO in every 1-2 weeks to discuss the progress of the development. At the end of this phase, there were six good and solid product concepts, including FastCaf\'{e}. The team looked at the feasibility of each concept in order to decide which concept would be turned into a product, both technically and financially. As the result, in October 2013 the team decided to focus on the development of FastCaf\'{e}.
\begin{displayquote}
\textit{``So it is like the other ones, ... payment renovation space, it was still a very good idea but it is [a] very large concept even if you spent much harder on MVP like a really small MVP, you would have been holding up a product as you had 10 different MVPs to test it ... before you launched the official product, it would have been 6 to 10 months investment before you saw the revenue.''} -- Team Lead
\end{displayquote} 

\subsubsection{Steering Phase}
In developing FastCaf\'{e}, the team did not employ Kanban as it was used in the development of other products, but Scrum. Customer feedback was collected through face-to-face interview with the caf\'{e} owners. Based on the feedback, the business analyst wrote the user stories and put them into the feature backlog. The team used the Fibonacci approach to estimate the effort and size. Then, together with the product manager and the design team, they prioritised the features that would go into each sprint.

In the envisioning phase, the backend side was performed by the team physically, who would go to the caf\'{e} and place the actual order and pay on behalf of the customers. In this phase, the team prioritised designing and building a new way to handle ordering and payment processing.
\begin{displayquote}
\textit{``So what happen when the order comes in? How is the payment process? How does that fit in with the flow, their work flow? When you go from a caf\'{e} to a very broad [rollout], you look at things at the beginning you think it is all the same. But it is completely different for each one.''} -- UX Designer Lead
\end{displayquote}

To solve the problem, the team collaborated with the caf\'{e} owners to co-create processes that fit seamlessly into their current processes and made sure that the technology was subsumed into the background. It was still unclear what the business model would be. Hence, at the first launch in February 2014, the team did a pilot program for one month to see how the payment system would work. The pilot program collaborated with 20 caf\'{e}s. It was extended for another three months to do a stress test of the system.
\begin{displayquote}
\textit{``Our strategy was to make [FastCaf\'{e}] free for venues who wanted to join our trial because we wanted to build and discuss so we needed that, also it gave us a delay by when things went wrong, because they were not paying for, they give us a bit bigger space [to improve]. If they are signed up, if they are paying that, we would be in trouble."} -- Team Lead
\end{displayquote}

It took 10 weeks after the launch for the team to get the first revenues from the customers. Only a few customers signed up to FastCaf\'{e} and started paying after the pilot program. They only made 20--30 dollars a week. The team learned that customers were willing to explore a new way of payment if it was convenient. For example, customers asked for PayPal as the way to handle of the online payment. However, PayPal charged 4 dollars per transaction, which was not financially viable. Hence, the team had to find a new way to secure the payment process.
\begin{displayquote}
\textit{``We did not have a payment system setup yet. ... There is something huge missing there, which is the barrier to get people to put in their credit card detail basically. ''} -- UX Designer Lead
\end{displayquote}

The team met the CEO every six weeks during this phase to report the findings and discuss the plan for the next six weeks. The CEO was more concerned about the number of caf\'{e}'s whilst the team looked at the number of orders and the amount of revenue.

After the pilot program, the team started charging the caf\'{e}s if they kept using FastCaf\'{e}. The business model was a subscriptions model, which was based on the volume of transactions. The caf\'{e} would pay a weekly fee and a percentage of the revenue earned through FastCaf\'{e}. For the electronic payment system, FastCaf\'{e} used electronic funds transfer at point of sale (EFTPOS). Hence, the payment from the customer was directly deposited to the caf\'{e} 's bank account. \\

\subsubsection{Acceleration Phase} 
\label{sec:market_creation_phase}
In this phase, the team focused on promoting FastCaf\'{e} to both potential customers and caf\'{e}s
to increase the revenue growth. The team went to a large scale event with a large number of people to get them to purchase or sign up. When the customers came down to pick up the coffee, the team interviewed about their experience with FastCaf\'{e}.

On the development side, using a continuous deployment approach the team kept improving the integration of FastCaf\'{e} with caf\'{e}s' work flow and also developed the payment system. There was no pivot after the envisioning phase.
\begin{displayquote}
\textit{``There was [a] suggestion of [pivoting]. Not necessarily pivot but [extending it] in further. I always believe that there was a broader ecosystem that we needed to tap in to have more control and access. ... Not just processing the payment, actually controlling the payment and going into their accounting software basically. I am looking into that. It would not be a pivot but extension of the offering.''} -- UX Designer Lead
\end{displayquote}

FastCaf\'{e} has generated 10\% of the expenses of the project. The biggest expense of the project was the salary since the employment cost is high in that country. However, the product attracts many users from other sectors.
\begin{displayquote}
\textit{``[The customers] kept asking us, `Can I use it for my florist?'. So, there might be something [to] work [on]. We give it a go.''} -- Team Lead
\end{displayquote}

\subsection{SeeSay}
\begin{figure*}[htbp]
    \centering
    \includegraphics[width=\textwidth]{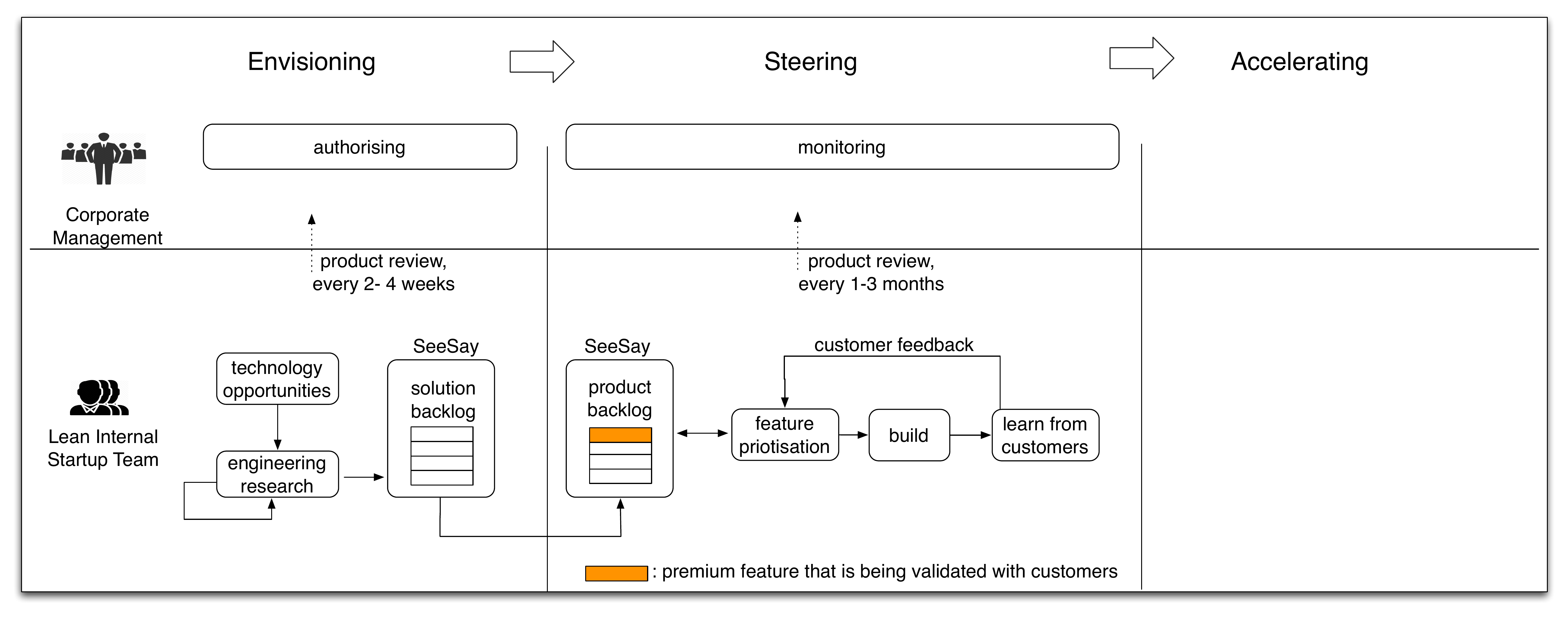}
    \caption{The Lean internal startup process of SeeSay}
    \label{fig:case_b}
\end{figure*}

The key Lean internal startup process of SeeSay is presented in Fig. \ref{fig:case_b}. The key metrics that were collected during each phase are presented in Table \ref{tab:measures_seesay}.

\subsubsection{Envisioning Phase}
For years, CallTech has been used to outsourcing any software development to external companies. Later on however, it was increasingly considered important to have internal software development in order to seek new opportunities from the existing technology. It was the Team Lead's vision to develop a new audio and video communication tool, when he was working on a particular project. The existing video conference solutions were not able to solve their problems.
\begin{displayquote}
\textit{``[The team] stumbled upon a technology, and they were sick and tired of video conferencing never working. So they thought to make their own. Based on WebRTC (Web Real Time Communication, a new technology that was a standard in Chrome and Firefox in 2013), this had to be really simple. They did it, and they thought `let's make this available for everyone'''} -- Chief Innovation Officer
\end{displayquote}

Three internship students were recruited to work on this project. In 2013, the first MVP was released and demoed internally. When the internship period was over, the development was taken over by a group of internal engineers. The team spent a long time establishing a solid team. At the beginning, they got to know each other and find their role in the team. Everyone had a lot of ideas and wanted their idea to win. Hence, most of the time, they were discussing and figuring out what should be in the product. Once agreement had been reached about the roles, the team started being more productive.
\begin{displayquote}
\textit{``If you look at the GitHub commit log, you can see that we were not so productive at the start. We did not have a lot of commits. At the end, one agrees on some kind of role distribution internally in the team: who does what, and who is an authority on the different fields. People start settling into their roles, and then the team starts being more performing ... When everyone agrees on the road ahead, ... then we start to be productive... We spent a lot of time that first fall to figure out what this product was going to be, what were we to create... We have used a lot of time to find a process where we involve people and let them contribute to designing features and designing the product.''} -- Co-Founder
\end{displayquote}

During the envisioning phase, the team only focused on how the technology could solve the problem. They did not find a way to monetise the product. Even though they were familiar with Lean Canvas, they did not use it at this stage.
\begin{displayquote}
\textit{``For me the Lean Canvas shows signs of demanding you to have all the answers right away. That is at least how I have seen it being used. Before you can start the project, you have to fill in the entire canvas, [but] you do not always have all the answers to everything in an early phase, for example how to earn money. I am certain that you do not have the answer to that in an early project phase. Several years can pass before you have the answer to that. That was also the case for us. I am certain that if we were to be evaluated based upon a Lean Canvas exercise early in the project, then this project would not get started.''} -- Team Lead
\end{displayquote}

At CallTech, each product was reviewed by the VP Product Management and Innovation Group. The team did not collect any metrics related to users in this phase. Instead, they were reviewed by the learnings they had based on the results of the engineering research about the product.

\begin{table}[htbp]                                                                 
\caption{Measures collected by SeeSay}                                             
\label{tab:measures_seesay}                                                            
\begin{tabular}{|p{2.5cm}|p{5.5cm}|p{4.5cm}|}                                        
	\hline                                                                   
	Phase & Measure & Cause-and-effect  \\                                          
	\hline                                                                          
	Envisioning & Number of tests on different types of product, number of product prototypes & Technology\\ \hline
	 \multirow{ 2}{*}{Steering} & Number of users, usage per day, per week & User behaviour \\ \cline{2-3}                             
	 &  User growth  & Business model \\                      
	\hline                                                                    
\end{tabular}                                                                    
\end{table} 

In the case of SeeSay, the role of an organisational champion was not recognised in any phase. The team had to find their own members to work on the project. Internship students were recruited at the beginning phase. 

The decision to move to the next phase was made during the product reviews. If the result of the product review in the envisioning phase had been poor, the Chief Innovation Officer would have been responsible for terminating the project. Instead during the steering phase, the decisions were made by the CEO.

\subsubsection{Steering Phase}
In the steering phases, the team still focused on validating their product in the market, whether the product attracted new users or not, instead of on financial return.
\begin{displayquote}
\textit{``[Our goal is] to get a lot of users, and building opposition is really a result, not one in the money''} -- Team Lead
\end{displayquote}

The team had iterated on their process to find the one that worked for the team in their current setting. The process was reviewed at the end of a two-week sprint, as inspired by Scrum. The team started collecting their feedback.
\begin{displayquote}
\textit{``The user's habits change over time. If we do not continuously improve our product, the users will change their habits, and stop using it. To us it is very important to invest a lot in research about the users' needs, habits, and behaviour, to ensure that our product is solving a real problem.''} -- VP Product Management and Innovation
\end{displayquote}

All the users' feedback was put into the product backlog and the Team Lead prioritised and decided which stories would be implemented. The tasks were assigned by considering their scopes. However, each feature was implemented by the same person, from end-to-end. The team used JavaScript both for the back- and the front-ends. Every new feature was launched to the service and tested through an experimentation with real users. The team also evaluated existing features that had been in the service. Features that were no longer working or those used least by users were removed from the service.
\begin{displayquote}
\textit{``When we have implemented something, then that functionality is launched into the service. Then we ask the users, and look at the data on what effect this has on the service. That can be done in different ways. [For example] we launch the functionality for half of the user mass, to see if it gives any effect towards some set goal. If not, then that experiment ends. But if it works, then we can launch it for
all the users.'' } -- Team Lead
\end{displayquote}

Instead of working from start-to-finish, CallTech set timeframes between product reviews. In the product review, they presented what they had done, what hypotheses were looked into, what they had learned and what they needed for the next round of development. They cooperated on setting goals for what should be achieved until the next product review. The product review was done differently depending on the phases.
\begin{displayquote} 
\textit{``The purpose is to know `OK, what did you learn since last time? What are the new risks? What problem is to be solved now?' Then we consider month by month. Then we meet again after 3 months, and do a new evaluation. Where are we now? Did we manage to solve those problems? What are the new problems? Is it worth it to keep going, or did something come up that causes the show to stop?''} -- Chief Innovation Officer
\end{displayquote}

At the moment, the team is developing a premium feature, a paid version of SeeSay. The co-founder is taking the responsibility to lead the development of premium features. At the same time, the size of the team is getting bigger. More new members with sales and marketing backgrounds are joining the team to create a market for the new features.

\subsection{Summary of Lean Internal Startup Process}
Looking back to our Lean-ICV framework, the summary of key activities in FastCaf'{e} case is presented in Table \ref{tab:key_activities}. The symbol \ding{54} means that the corresponding actor is not found the case, thus no activity is identified. In the FastCaf\'{e} case, the B-M-L loop was used to identify the potential problem and solution. The team used different methods to reveal customers' need. In the case of SeeSay, as the product was highly driven by the technology, during the visioning phase, the team aimed to turn the idea into real product.

\begin{table*}[htbp]
	\centering
	\caption{Key activities in the two cases}
	\label{tab:key_activities}
	\begin{tabular}{|p{2.2cm}|p{3.8cm}|p{2.9cm}|p{2.9cm}|}
	\hline
	 Phase & Actors & FastCaf\'{e} & SeeSay \\ \hline
	 \multirow{3}{*}{\parbox{3cm}{Envisioning}} & Corporate management  & Authorising & Authorising  \\ \cline{2-4}
	 & Lean internal startup team & Solution and business identification & Technology research \\ \cline{2-4}
	 & NVD/External consultant & Coaching \& training & \ding{54} \\ \hline 
	  \multirow{3}{*}{\parbox{3cm}{Steering}} & Corporate management & Monitoring & Monitoring  \\ \cline{2-4}
	 & Lean internal startup team & Solution and business validation & Solution validation  \\ \cline{2-4}
	 & NVD/External consultant & Coaching \& training & \ding{54} \\ \hline 
	 \multirow{3}{*}{\parbox{3cm}{Accelerating}} & Corporate management & Monitoring & \ding{54} \\ \cline{2-4}
	 & Lean internal startup team & Business scaling & \ding{54}   \\ \cline{2-4}
	 & NVD/External consultant  & \ding{54}  & \ding{54}  \\ \hline 
	 \end{tabular}
\end{table*}

In the case of FastCaf\'{e}, the selecting, rationalising and delineating activities were not recognised in this stage. One of the reasons is the selecting and rationalising activities already happened in the middle of the Steering phase, and it was not carried out in order to evaluate the startup's performance, but rather as the consequences of the new management's policy. In this situation, the organisational championship is an important mediating role to solve this issue. The second reason is that top management had been involved in the Lean internal startup process since its conception. Before the actual development took place, the Lean internal startup team were required to present their ideas in order to get authorisation from the top management. This made all their activities and progresses transparent to the top management. 
\section{The Enablers and Inhibitors for Lean Internal Startups}
\label{sec:comparison}
In this section, we identify the enablers and inhibitors for Lean internal startups in both cases studied. 

\subsection{The Enablers for Lean Internal Startups}
Table \ref{tab:enablers} summarises the key enablers for Lean internal startups and their outcomes in all cases. The \ding{52} symbol means that the factor is found in the corresponding case. An empty cell means that the factor is not found in the corresponding case. The enabling factors are grouped in sub-category and category, which are defined in the Conceptual Framework as described in Section \ref{sec:theoretical_framework}. Table \ref{tab:inhibitors} should be read in the same way.

CallBook consistently focused on the innovation initiative. FastCaf\'{e} was considered as a growth strategy exercise to diversify the company's portfolio. The company hired an external consultant to support the initiative. Even though there was a change in the strategy due to the change in the company ownership, the initiative continued.

The consistent focus on innovation was maintained because the initiative was supported by top management, which in this case was the CEO. The CEO was the one who protected the initiative when a new management came in. The CEO also secured all the resources needed by the team. Thus, the team was able to focus on product and business development. The support from top management at the time it was needed inspired the team and improved the confidence level of the team.
\begin{displayquote}
\textit{``[The support from the CEO] was hugely important. It gave us to believe that we are unto something, pretty special.''} -- UX Designer Lead, FastCaf\'{e}
\end{displayquote}

In the SeeSay case, the teams used resources that were available in the company. They were not required to figure out how to generate financial return as soon as possible, as in a standalone startup, but instead could focus on the quality of the product and the customer satisfaction.
\begin{displayquote}
\textit{``We do not have to think about [where to get money from]. We can use our resources to continue developing our product. ... We have used that [resources] on scaling up in our number of users. So when we are going to start earning money, then we already have a large platform to stand on, with a lot of users using our service. ... We feel very lucky to have that.''} -- Team Lead, SeeSay
\end{displayquote}

\begin{landscape}
\begin{table*}[htbp]
	\centering
	\caption{Enabling factors for Lean internal startups}
	\label{tab:enablers}
	\begin{tabular}{|p{2.5cm}|p{2.2cm}|p{4cm}|p{7.5cm}|p{1.5cm}|p{1.1cm}|}
	\hline
	Category & Sub-category & Factors & Identified outcome & FastCaf\'{e} & SeeSay \\ \hline
	\multirow{5}{*}{\parbox{1.8cm}{Organisational Structure}} & Strategy &Explicit strategy on innovation & Justification for the existence of the internal startup in the company & \ding{52} &  \\ \cline{2-6}
	 & \multirow{2}{*}{Leadership} & Top management support  & Secured the budget and resources & \ding{52} & \ding{52} \\ \cline{3-6}
	 & & Permission to break the rules & Speed up the development process, improved the current practices in the company & \ding{52} & \\ \cline{2-6}
	 & Champion & Organisational champion & Protection from strategic change & \ding{52} &   \\ \cline{2-6}
	 & \multirow{2}{*}{Resources} & Company's brand and reputation & Access to existing customers & \ding{52} & \ding{52} \\ \cline{3-6}
	 &  & Branch offices and departments &  Access to existing network of experts in different areas within the company &  \ding{52} & \ding{52} \\ \hline
	 Knowledge and Technology & Knowledge Management & Coaching, mentoring and training & Built team confidence and improved the skills of the team members & \ding{52} & \\ \hline
	Culture &  \multirow{3}{*}{\parbox{1.6cm}{Empower-ment}} & Autonomy in decision-making process & Speed up the development process, improved learning process & \ding{52} & \ding{52}    \\ \cline{3-6}
	& & Personal stake in the outcome & Improved motivation of the team & \ding{52} &   \\ \cline{2-6}
	 & Trust & Freedom to experiment and pivot & Improved learning process & \ding{52} & \ding{52}\\ \hline
	\multirow{2}{*}{\parbox{1.8cm}{Human Resources}} & Internal collaboration & Cross--functional team & Increased collaboration and reduced communication overhead & \ding{52} & \ding{52} \\ \hline      
	 
\end{tabular}
\end{table*}
\end{landscape}

As part of a large company, each of the employees had a specific KPI as the basis for a pay raise or bonus. There was no additional reward given to the FastCaf\'{e} team. It was the intrinsic reward that the team got while working on the internal startup.
\begin{displayquote}
\textit{``It is just the motive that we built [a] cool product that is changing people lives.''} -- Team Lead, FastCaf\'{e}
\end{displayquote}

Both the Team Lead and the UX Designer Lead of FastCaf\'{e} recognised that the training in Design Thinking was helpful for them to establish a new idea. Desirability, viability and feasibility became the main lens for finding and validating a good idea and a product concept.
\begin{displayquote}
\textit{``[Design Thinking] really helps to  find market fit at [a] very nascent stage to get a proof of concept any way.''} -- UX Designer Lead, FastCaf\'{e}
\end{displayquote}

The decision to join the team was driven by intrinsic motivation rather than extrinsic motivation. Gaining new experience was the biggest motivation for getting involved in this innovation initiative. Some of the members had to be demoted from their positions before joining the team but working together with the IDEO was deemed a good opportunity.
\begin{displayquote}
\textit{``So this company is giving me to do something I never could have done before and it is really motivating ... I do not need extra money ... While [for] people in [a] startup, the funding, the motivator is on that I will make a lot of money on this. But I am not making money on [FastCaf\'{e}] but I am getting incredible experience.''} -- Team Lead, FastCaf\'{e}
\end{displayquote}

In the case of SeeSay, to achieve this goal, the team has developed their own processes, which was easier than following the routines. The team also used tools available on the Internet, instead of making them themselves. Furthermore, the team also had the competences needed in the team, which allowed them to test things earlier before finding on actual solution.
\begin{displayquote}
\textit{``It is important that the team includes both the one doing the user research, and having worked with user needs, and a developer that can understand what is possible, and a designer to lead the creative part, and figure out what we actually can do, and test the design.''} -- Team Lead, SeeSay
\end{displayquote}

\subsection{The Inhibitors for Lean Internal Startups}
Besides the key enablers that foster Lean internal startup in large companies, internal startup initiatives also suffers from inhibitors that affect its performance. Table \ref{tab:inhibitors} summarises the key inhibitors for Lean internal startups.

As a consequence of the size and complexity of the modern business, large companies tend to be bureaucratic, which lowers the companies' agility to be innovative. This is also the case at CallTech.
\begin{displayquote}
\textit{``There are a lot of policies in a large company, which are created for a large company, and not a small team. They are policies on everything from procurement, contracts, and such, where there are rules for how things should be done, and they do not always fit us. They can easily get in our way. ... To start innovation in a large company, it has to be done in such a way that you avoid the policies that apply for the large company applying for the people doing innovation.''} -- TeamLead, SeeSay
\end{displayquote}

All interviewees considered CallTech as an traditional and bureaucratic telecommunication company. The employees were measured on a quarterly basis as the company is measured in the stock market. An innovation project spanning between 3-5 or even 5-7 years, might not be in the interest of the company. Moreover, for SeeSay, there is no incentive to get the project succeed. Hence, being part of an internal startup is not of interest for the employees. Moreover, CallTeach does not have a strong brand among software people in the job market. That is why the team hired intern students to start the product development.

The team had autonomy to decide on the business model for the new product. This raised a tension among the employees within the company.
\begin{displayquote}
\textit{``When you are part of the innovation team and you get pulled away to do a special project, you can get a bit resistance sometime from those who are left behind or left on the corner to do day-by- day stuff.''} -- UX Designer, FastCaf\'{e}
\end{displayquote}

As part of a large company, the internal startup team has access to various types of expertise inside and outside the company to build the product. For example, they could use the technology or platform that was developed by another team in the company. If this happened, the internal startup team would easily become dependent on others, who might not prioritise them in their continued development. As the result, the team could easily lose its focus on customers.
\begin{displayquote}
\textit{``It is easily done in a large company that you get promised something from some other internal team. They say, `We will fix this for you. We will make it, just trust us.' They maybe mean it seriously, but their priorities can change fast. If what you need takes half a year, then maybe after 2 months they say, `No, we need to make this instead'. You have already lost a lot of time because you trusted them to do it.''} -- Team Lead, SeeSay
\end{displayquote}

As suggested by the Lean startup approach, pivot is common to any startup to avoid bankruptcy \citep{ries11}. This is not the case in an internal startup. Even though the decision to pivot is made in the team, but they need to gain an approval in order to continue the process. Otherwise, pivoting will lead to a termination.
\begin{displayquote}
\textit{``But I think you need a kind of compensation about funding and resources if you decide to pivot. It really depends on the funding, like who is paying for it and how is it being paid for?''} -- Team Lead, FastCaf\'{e}
\end{displayquote}

In the accelerating phase, the team now already has paying customers. At the same time they also have to continue the development based on the long-term plan. This causes another challenge for the team keeping their eyes on the long term goal without getting distracted by short term issues.
\begin{displayquote}
\textit{``It is really easy to do small fixes for things that are right for your paying customers. But your long term goals [are things like] recognising the revenue. They are much more complex [than small fixes]. So you really have to [focus on] that. When a small [fix] happens that makes a lot of noise. You have to be really careful. I know the problem but I am not  fixing it now, it is really hard for customers, because [the customers] are paying me. It is really important for them, but it might be not a chase [for long term goals] and it is the hardest thing.''} -- Team Lead, FastCaf\'{e}
\end{displayquote}

\begin{landscape}
\begin{table*}[htbp]
	\centering
	\caption{Inhibiting factors for Lean internal startups}
	\label{tab:inhibitors}
	\begin{tabular}{|p{2.5cm}|p{2.2cm}|p{4cm}|p{7.5cm}|p{1.5cm}|p{1.1cm}|}
	\hline
	Category & Sub-category & Factors & Identified outcome & FastCaf\'{e} & SeeSay \\ \hline
	\multirow{4}{*}{\parbox{1.8cm}{Organisational Structure}} & Management & Policies and guidelines & Slowed the development process down & & \ding{52} \\ \cline{2-6}
	 & Strategy & Changes in corporate strategy & Could lead to termination of the initiative & \ding{52} & \\ \cline{2-6}
	 & Leadership & Permission to break the rules & Raised a potential internal conflict in the company & \ding{52} &   \\ \hline
	 Knowledge and Technology & Technology & Reliance on technology or platform developed by other teams (internally or externally) & Lost focus on customers, highly depended on other teams & & \ding{52} \\ \hline
	\multirow{2}{*}{Culture} & Trust & Lack of freedom to experiment and pivot & Limited the learning process & \ding{52} &  \\ \cline{2-6}
	& Empower-ment & No personal stake in the outcome & Discouraged motivation to innovate&  & \ding{52} \\ \hline
	\multirow{2}{*}{\parbox{1.8cm}{Human Resources}}& Human capital & Job description, routines & Difficulties in recruiting and building the team & & \ding{52}   \\ \hline
	\multirow{2}{*}{\parbox{1.8cm}{Business Characteristics}} & Customer orientation & Balancing the long--term vs. short--term issues & Dilemma to satisfy current customers vs. focusing on long--term goals & \ding{52} & \\ \hline
\end{tabular}
\end{table*}
\end{landscape}

\section{Discussion}
\label{sec:discussion}
This section discusses and makes sense of the findings of the multiple-case study. It is composed of three main parts. The first two parts will revisit the research questions of this study as listed in Section \ref{sec:introduction}. The third part will discuss the limitations of this study.

\subsection{RQ1 - Lean Internal Startup Processes}
The first principles of the Lean startup approach suggests that anyone can be an entrepreneur without owning a business, for example a student or an employee within a corporation. In both cases, the team leads of the founders were the employees in a middle management position. With this unique role, middle managers link and reconcile top management's strategic direction with implementation issues surfacing at the operational level \citep{kanter82,wooldridge08, glaser16}. 

\begin{figure*}[htbp]
    \centering
    \includegraphics[width=1.0\textwidth]{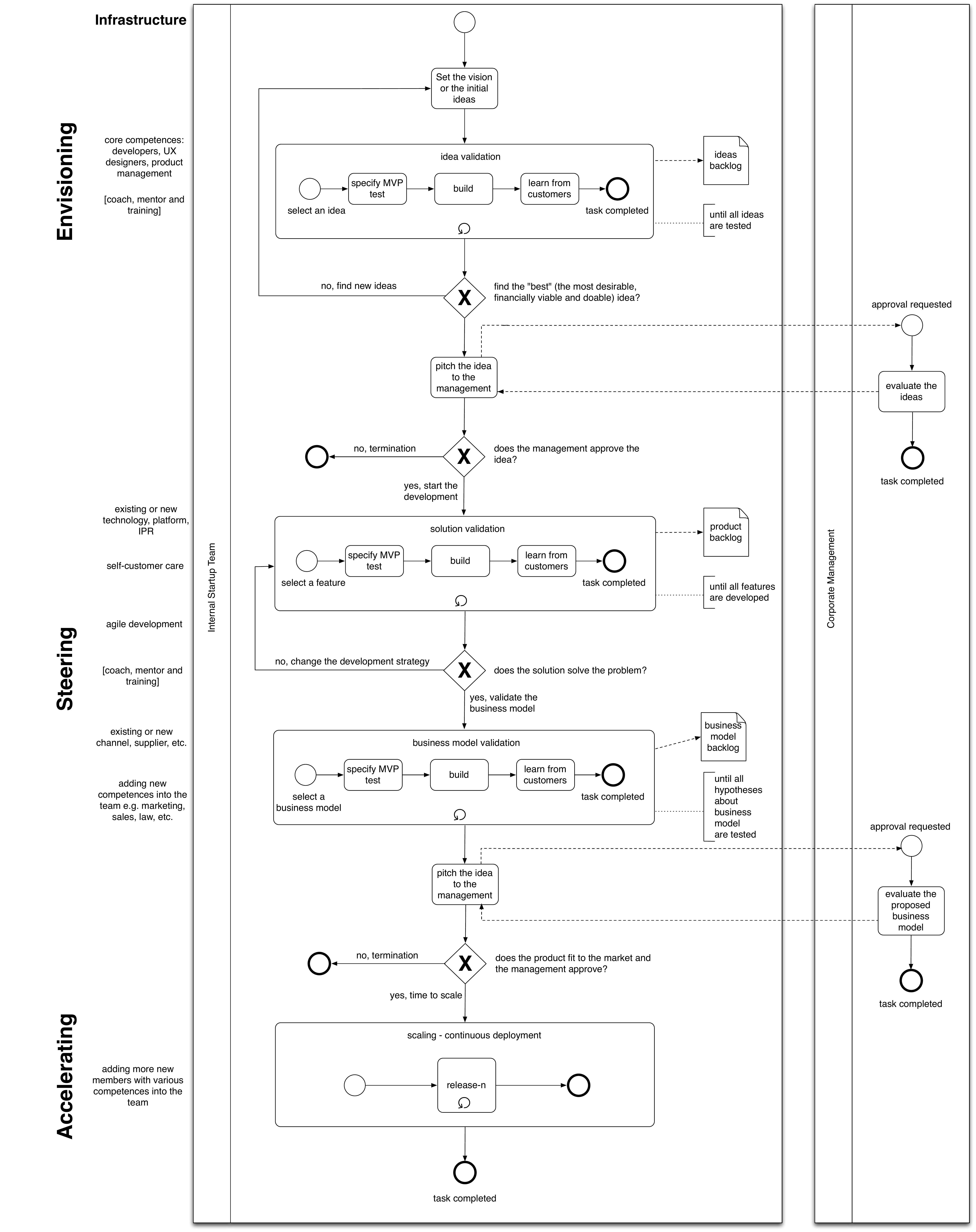}
    \caption{The Lean internal startup process}
    \label{fig:lis}
\end{figure*}

The general Lean internal startup process is illustrated in Fig. \ref{fig:lis}. The infrastructure needed for each process is shown on the left side of the figure. The text inside square brackets means that the findings are found only in one case. We used Business Process Model and Notation (BPMN) to show the general process. The first lane shows the processes performed by internal startup, while the second lane by corporate management.

In the envisioning phase, our findings show that Lean internal startups that are initiated by corporate management has no founder with a vision for the new product. Instead of translating the vision into business hypotheses as suggested by Lean startup approach, the internal startup looks for interesting user problem to solve. Unfortunately, Lean startup approach does not have explicit method or technique in the ideation process \citep{mueller12}. As shown in the case of FastCaf\'{e}, Design thinking approach can complement Lean startup approach in this ideation process.

Design thinking is an user-centred approach to generate innovative solutions for wicked problems \citep{buchanan92,thoring11}. Design thinking approach is not an intuitive and individual process, but rather proposes different process steps an ideation techniques. Unlike typical creative design process, which is individualistic, the idea of design thinking is to be applied by multi-disciplinary team, instead of well-trained designers.

Design thinking consists of six processes: understand, observe, point of view, ideation, prototyping and test. Unlike Lean startup approach, which starts with a business idea, design thinking approach starts with a problem and question. The idea is developed within the fourth process, ideation. To generate this with this idea, there are secondary research (understand) and user research (observe). The knowledge gathered from these research is then condensed into a point of view, which describes a micro theory about user needs. Based on this, innovative ideas are generated to solve the needs. The selected idea is then prototyped and tested to the market to get feedback from prospective users.

In the course of steering phase, the Lean internal startups grow into a specific one-product business. Both cases suggest that once an idea has been confirmed and approved by management, there is no way for the internal startups to pivot back to the envisioning phase. The final idea, that is presented to management, is the ``best'' idea, in terms of desirability, financially viability and feasibility. This means that the idea is the one most desired by the customers, shows high potential growth or revenue and can be implemented within the company.

The steering phase has two sequential activities: solution validation and business model validation. In the solution validation, by using agile methods, the team implements all the key features that have been identified in the previous phase. To achieve the problem/solution fit, the team constantly communicates with the customers. New features are released to the market as frequently and within short intervals. In this phase, communication with the customers is managed by team itself. Any feedback from the customers is received directly by the team. Once the product has achieved problem/product fit, the team validated the business model that had been identified earlier. The team did not have fully freedom to experiment on new business model, but rather finding a way to scale the business.

The typical metrics collected and maintained by the Lean internal startups are related to the usage of the product, e.g. number of users, number of activation, etc. Corporate management is more interested to look at the metrics related to the objective of the team. In the case of FastCaf\'{e}, the objective was to generate revenue; thus the corporate management focuses on the metrics like number of revenue, number of orders. In the case of SeeSay, the collected metrics were typically used for the team internally. The product was reviewed by the corporate management in terms of the learning they had during the course.

In the accelerating phase, our case study finds that the internal startups are in a stable stage. The internal startups are running like an established company now. They already have a product in the market, a business model and paying customers. In this stage, the objective is to scale and generate more revenue, but at the same time they have to satisfy the specific needs of paying customers. This classic dilemma is not only for internal startups also for established and large companies in general: to increase market share and maintain customer's trust \citep{ford10}. In this stage, companies tend to pursue incremental innovation, which delivers minor changes in the product and minor customer benefits \citep{chandy98, mcmillan10}. 

\subsection{RQ2 - Enablers and Inhibitors for Lean Internal Startups}
Our case study also identifies the enablers and inhibitors of applying Lean startups in large companies. The common key enablers for Lean internal startup are top management support and cross-functional team. With the top management support in place, the Lean internal startup teams secure the budget and resources they needed. In all of the cases, the Lean internal startup initiative did not operate, under a specific department or division, but rather in a cross-department setup. The teams directly report to the top management. Hence, whenever the teams require anything, they must submit the request to the top management. To get this support, our cases suggest that the Lean internal startup teams need to convince the top management that they are working on the best idea, which will bring revenue to the company and will not potentially disturb the existing business or customer-supplier relationship.

Cross-functional team means that the team consists of the members with various backgrounds and roles, e.g. software development, marketing, sales, etc. This configuration is needed to improve the decision-marking process, increase collaboration and reduce communication overhead. To validate all the hypotheses about the new product, the team needs to speed up the development process and test to the market iteratively. Our cases suggests that in the early phase, the teams mainly consists of members with a software development background. As the product grows mature, members from marketing and sales are recruited to grow the business.

As shown in Table \ref{tab:inhibitors}, our case study finds that each Lean internal startup deals with different challenges. One of the reasons is that FastCaf\'{e} and Seesay had different inception process.  FastCaf\'{e} was driven by company's strategy to increase the revenue, whilst SeeSay was employee-driven innovation initiative. Hence, in the case of FastCa\'{e}, the biggest challenge was the strategic change. On the other, in the case of SeeSay, the founder had to find a way to avoid policies and guidelines that would slow their process down. In such situation, the FastCaf\'{e} team was backed up by the CEO, but in the SeeSay case, it was the founder's job to solve the issue.

The second reason is related to the objectives of the Lean internal startups. In the case of FastCaf\'{e}, the initiative was aimed at revenue. Therefore, since the envisioning phase, the team had identified a potential business idea. However, once the business idea had been approved, there was no chance to pivot. In the case of SeeSay, there was no such requirement. Hence, the team focused on the quality of the product.

Another reason was related to the type of product developed by both teams. In the case of FastCaf\'{e}, the development of back-end application had to take into account the caf\'{e} payment system. Hence, the team had to balance the long-term goal, which was to grow the business and short-term goal, which was to satisfy the current customers. In the case of SeeSay, the product was mainly driven by the company. The product was managed only by the team and there was no third system that should be served by the product.

In the case of FastCaf\'{e}, the selecting, rationalising and delineating activities were not recognised in this stage. One of the reasons is the selecting and rationalising activities already happened in the middle of the Steering phase, and it was not carried out in order to evaluate the startup's performance, but rather as the consequences of the new management's policy. In this situation, the organisational championship is an important mediating role to solve this issue. The second reason is that top management had been involved in the Lean internal startup process since its conception. Before the actual development took place, the Lean internal startup team were required to present their ideas in order to get authorisation from the top management. This made all their activities and progresses transparent to the top management.

To look deeper on the Lean internal startup processes, we used the Lean-ICV framework. The framework allows us to identify the key processes in both product and business development. Our study results show that in the top-down approach, during the envisioning process, the corporate management are involved in defining the objective of them. This is different with the bottom-up approach, where the founder defines the goal of the initiative. In the acceleration phase, our study results also show that  the selecting, rationalising and delineating activities are not recognised. In both cases, the progress of the Lean internal startups is transparent to the corporate management, since the teams are reported and evaluated directly by them.

\subsection{Validity Threats}
Threats to validity \citep{runeson12,wohlin12} related to the results from this study have been identified and are discussed below.

\subsubsection{Construct Validity}
Construct validity refers to the extent to which the operational measures, e.g. the constructs discussed
in the interview questions that are studied, represent the objective of the study \citep{runeson12,wohlin12}. In this study, two pilot interviews were conducted to validate whether the questions were interpreted in the same way by the researchers and the interviewees. As discussed in Section \ref{sec:collection_analysis}, the pilot interviews were also intended to test the conceptual framework. In addition, data source triangulation was used to strengthen the evidence generated in this study.

\subsubsection{Internal Validity}
The retrospective analysis nature of this study makes it vulnerable to historical types of internal validity threat \citep{runeson12,wohlin12}. This is related to historical events that occurred during the product development that may affect the accuracy of cause-effect relationship \citep{runeson12,wohlin12}. A company representative with extensive knowledge about product innovation helped with constructing a time-line of the development, identifying the key processes as well as the success and the challenges faced by the team. Company documentation was used to capture detailed events and practises that occurred during the development process.

Another issue related to internal validity threat is the selection of participants. Due to geographic constraint and organisational changes, it was not possible to involve all core team members from each cases. The participants from the FastCaf\'{e} case were two out of 5 core members of the team. One participant from the FastCaf\'{e} was still employed at the time of investigation (Team Lead) but the other one had left the company for one year (UX Designer Lead). The rest of the core team members were not employed in the company anymore. We were unable to collect more data from the rest of the members. However, we believe the this issue was mitigated by the characteristics of our interviewees. First, both interviewees had extensive knowledge about FastCaf\'{e} through their involvement in the exploration, validation and creation the market. Second, the interviewees who helped this study covered diverse and crucial roles that were vital to the development of FastCaf\'{e}. In the case of SeeSay, all the interviewees were founders of the internal startups and members of the company's top management that worked closely with the internal startups. In addition, all of them are still employed.

The number of interviewees involved in this study could raise the threat related to the trustability of the interviewees' data. As we described in Section 4 that in the case of FastCaf\'{e}, we interviewed 2 key members of the internal startup in three rounds of data collection. This strategy allowed us to cross-check the information we got from the previous interview. In addition, to achieve data triangulation, we look at and review the available and relevant supporting materials internally from the company. We also collected materials from newspapers, magazines, and other published materials to balance the interview's data.

\subsubsection{External Validity} 
External validity is concerned with the extent of generalisability of a study \citep{runeson12,wohlin12}. The study presented here is a multiple case study from two different products from two different companies. However, providing a detailed description of the context of the products (see Section \ref{sec:case_companies} and Section \ref{sec:findings}) helps in improving the study's external validity \citep{petersen09}. Even though each company and product innovation are unique, analytical induction helps to determine the generalisability between cases \citep{wieringa13}. Hence, we provide an in-depth analysis of each case and carefully describe the context and provide clear insights of a particular context. The reason for doing this is to help practitioners and researchers to easily compare the studied context and their own. There may be similarities with other context, such as in the process of product innovation through internal startup initiatives. In such situation, practitioners may take into consideration the enablers and inhibitors that we identified into their own process. For researchers, the detailed description of the context would help them to compare and synthesise our findings with similar contexts and thus provide complete evidence that can be useful to practitioners.

\subsubsection{Reliability Validity}
This aspect of validity concerns with to the extent to which the data and the analysis are dependent on the specific researchers \citep{runeson12,wohlin12}. Another researcher who has more extensive knowledge and experience in case study research was also engaged in the review of the design and execution of the study. In addition, the interview transcripts used for the data analysis were sent back to the interviewees for their review. These practises helped to reduce research biases during data collection and analysis.
\section{Conclusion and Future Work}
\label{sec:conclusion}

There are three contributions of this study for research. Firstly, while most research on the Lean startup approach is centred on new and emerging software startups, there is a lack of empirical research examining its implementation in large companies. In addition, software product innovation processes are not well-captured in the literature \citep{covin15}. Hence, our study is one of the first attempts to fill these gaps.

The second contribution of this study is that it has explored the application of the method-in-action framework on Lean startup research and identified the relevant factors that affect its use in a specific context. The conceptual framework, as discussed in Section \ref{sec:theoretical_framework}, provides an alternative way to further research on the use of Lean startup approach in non-startup context.

The third contribution of this study is the general process of Lean internal startup, and the evidence of the enablers and inhibitors, which is both theory-informed and empirically grounded. The Lean internal startup approach provides a better understanding of the essential practices of successful software product innovation. For practice, our study results have shown that when the company strategy is in place, it becomes the main enabler for the success of Lean internal startup initiative. On the other hand, it can also be the main reason for terminating such initiative, despite its positive result. In the latter case, the existence of organisational championship plays a crucial role in protecting the initiatives.

The approach we took in this study is not impervious to limitations, that may affect the outcome of this study. The number of interviewees from both cases are not balanced. Due to geographical constraint and organisational changes, we were unable to get more interviewees. However, we did examine every available sources to achieve data triangulation. We have achieved what we set out to do by understanding more about Lean internal startup as a potential approach that facilitates software product innovation in large companies.

We envision three avenues for future research. The first is to extend this study by addressing the limitations of the research approach used in this study. Future study could involve more cases that are in different phases in order to identify the conditions that impede or foster the Lean startup approach in the context of large companies. Another extension of this study would be to validate the enablers and inhibitors for Lean internal startups in a broader population, for instance through a questionnaire. The second direction of future research could perform a quantitative study to investigate the impact of Lean internal startups on the success of software product innovation. A number of metrics have been suggested to measure the success of innovation in the context of large companies e.g. \% of revenue generated by the new product, number of patents, etc. Such study could help to establish the cause-effect relationship between Lean internal startups and software product innovation statistically. A third and final suggestion is that future research could focus on comparative study on Lean startup approach in large companies context and standalone startups.



\section*{References}
\bibliographystyle{elsarticle-harv} 
\bibliography{elsarticle-bib}






\end{document}